\documentclass[aps,prl,showpacs,superscriptaddress,twocolumn,floatfix]{revtex4}
\usepackage{mciteplus}
\usepackage{graphicx,color}
\usepackage{amsmath,amssymb}
\usepackage{dcolumn}   
\usepackage{bm}        
\def\beq{\begin{equation}}
\def\eeq{\end{equation}}
\def\bea{\begin{eqnarray}}
\def\eea{\end{eqnarray}}

\def\ra{\rangle}
\def\la{\langle}

\newcommand{\eref}[1]{Eq.~(\ref{#1})}%
\newcommand{\fref}[1]{Fig.~\ref{#1}} %

\begin{document}

\title{Spatial Structures and Giant Number Fluctuations in Models of Active Matter}

\author{Supravat Dey}
\affiliation{Department of Physics, Indian Institute of Technology, Bombay,
Powai, Mumbai-400 076, India}
\email{supravat@phy.iitb.ac.in}

\author{Dibyendu Das}
\affiliation{Department of Physics, Indian Institute of Technology,
Bombay, Powai, Mumbai-400 076, India}

\author{R. Rajesh}
\affiliation{Institute of Mathematical Sciences, CIT campus, Taramani, 
Chennai-600113, India}

\date{\today}

\begin{abstract}
The large scale fluctuations of the ordered state in active matter
systems are usually characterised by studying the ``giant number
fluctuations'' of particles in any finite volume, as compared to the
expectations from the central limit theorem. However, in 
ordering systems, the fluctuations in density ordering are often captured
through their structure functions deviating from Porod law.  In this
paper we study the relationship between giant number fluctuations and
structure functions, for different models of active matter as well as
other non-equilibrium systems. A unified picture emerges, with
different models falling in four distinct classes depending on the
nature of their structure functions. For one class, we show that
experimentalists may find Porod law violation, by measuring subleading
corrections to the number fluctuations.
\end{abstract}   


\maketitle

Active matter, collections of interacting self-propelled particles, are found
in many different contexts. Examples include
bird flocks \cite{ballerini1,*ballerini2}, bacterial
colonies \cite{ZhangPnas10}, actin filaments propelled by molecular
motors \cite{Volakar} and vibrated granular rods and disks
\cite{Sriram,kudrolli,ChateDisk}. In these, the ``activity'' refers to
conscious decision making or internally generated cellular thrusts in the
biological systems, or impulses from a vibrating plate for
the granular systems. The combination of activity and interaction can lead to
macroscopic order \cite{Vicsek,*Vicsekrev,TonarPrl98,ChateCoh,*ChatePre08,ChateAnPrl06,ChateSpr,Chatetopological}.
However, these systems are far from equilibrium and
the usual notions of equilibrium phase transitions come under unexpected
challenges \cite{TonarPrl95,TonarRev}. In particular,  
macroscopic order, and large scale fluctuations
reminiscent of critical equilibrium systems, coexist. 

Depending on their dynamics and symmetries, different active matter systems 
exhibit macroscopic polar, nematic, and/or density order. Polar
ordering has been demonstrated in point polar particle (PP) models
\cite{Vicsek,ChateCoh,*ChatePre08}, experiments with granular disks
\cite{ChateDisk}, and continuum theories
\cite{TonarPrl95,TonarRev}. For polar rods (PR), continuum theories
rule out macroscopic polar ordering
\cite{MarchettiPrl08,*MarchettiPre08}, and experiment on mobile
bacteria \cite{ZhangPnas10} and simulations of models of polar rods
are in agreement \cite{bar06,ChateSpr}. Apolar rods (AR) or active
nematics have been studied experimentally \cite{Sriram}, in
hydrodynamic theories \cite{SriramEPL03,TonarRev}, and in simulation
\cite{ChateAnPrl06} and exhibit nematic and density ordering. 

The density fluctuations in the ordered state has been characterised by the
number fluctuations $\sigma_{l}^2 = \la n^2 \ra_{l} - \la n\ra_{l}^2$ of
particles in a finite box of linear size $l$, where $n$ is the particle
number. In active matter systems, $\sigma_{l}^2 \sim \la n \ra^{\alpha}$ with
$\alpha > 1$, indicating ``giant'' number fluctuations (GNF) in comparison 
to what is expected from the central limit theorem. The exponent $\alpha$ has
been used to infer the long range correlations in the system.
In two dimensions, for the PP \cite{TonarRev,ChateCoh,*ChatePre08,ChateDisk}, 
and PR \cite{ChateSpr,ZhangPnas10} systems, it is now known that $\alpha =
1.6$, and for AR systems \cite{SriramEPL03,TonarRev,Sriram} $\alpha =
2.0$.

Consider now an active matter system relaxing to its ordered state
from an initial disordered state. As the density order grows with
time $t$, there is an increasing  macroscopic length scale $\mathcal{L}(t)$ 
over which there is enhanced clustering. Information about the 
spatial structures in such a coarsening system can be obtained by
studying the spatial density-density correlation function
$C({\bf r},t) = \la \rho({\bf 0},t) \rho({\bf r},t) \ra$ where
$\rho({\bf r},t)$ is the local density at point ${\bf r}$.
Systems relaxing to an equilibrium state typically exhibit 
clean domain formation \cite{Bray}, resulting in a linear form 
of $C({\bf r},t) = a - b|{\bf r}|/{\mathcal{L}(t)}$ for $|{\bf r}|/
{\mathcal{L}(t)} \ll 1$, known as the Porod law \cite{Porod}. On the other
hand, many systems relaxing to an nonequilibrium steady state violate Porod
law due to a hierarchy of cluster sizes.  Examples include
sliding particles on fluctuating interfaces 
\cite{DibyenduPrl00,*DibyenduPre01,*Manoj} and freely cooling granular 
gases \cite{MahendraPrl,*MahendraPre11}.

In this paper, we ask the following. First, we ask whether coarsening active
matter systems obey Porod law.
A few studies have addressed this question --
discrete models of active nematics \cite{SriramPrl06,*Chateunpub}, and
a recent numerical implementation of a hydrodynamic polar model
\cite{MarchettiPre2010} have shown non-Porod behaviour.
A further systematic study is necessary, and in this paper
we show that the Porod law is indeed violated by all the models that we study. 

Second, we ask whether the fluctuations that contribute to GNF are the 
same as those that cause Porod law to be violated. In particular, we ask
whether the large distance behaviour of $C({\bf r},t)$ can be deduced by
knowing $\alpha$. In general,
$C({\bf r},t)$ contains more
information than $\sigma_{l}^2$, as the latter is derived from the
former:
\beq
\sigma_{l}^2(t) = l^d \int_0^{l} d^d{\bf r} [C({\bf r},t) - \la
\rho \ra^2],
\label{eq1} 
\eeq 
where $\la \rho\ra$ is the mean density.
If the integrand decays to zero over
a length scale $\xi \ll l$, then $\sigma_{l}^2(t) \sim l^d \sim
\la n \ra$ for large $l$, or $\alpha=1$.  Since $\alpha>1$
for active matter, the upper limit of \eref{eq1} should contribute to the
integral, implying that non-trivial correlations extend beyond
the scale $\mathcal{L}(t) \geq l$.
Hence, one would expect the
behaviour of $C({\bf r},t)$ near $|{\bf r}|/\mathcal{L}(t) \approx 1$
to contribute to $\sigma_{l}^2(t)$ in \eref{eq1}, but we will see
below many interesting exceptions to this.
In this paper, we show that different active matter systems as well as
other nonequilibrium systems studied in other contexts,
fall is four distinct classes based on the relation between 
their $\sigma_{l}^2(t)$ and $C({\bf r},t)$. In case of the first type,
the small $|{\bf r}|/\mathcal{L}(t) \ll 1$ behaviour of 
$C({\bf r},t)$ has no bearing on the exponent $\alpha$. For the other three
types, it does, albeit in three distinct ways. 
   
{\it Type 1}: We start with PP systems. We first study numerically the
Vicsek model \cite{Vicsek}, which we denote as PP(1), in two
dimensions.  All particles move with constant speed $v_0$. The
positions ${\bf r}_i$ and velocity orientations $\theta_i$ of particle
$i$ at time $t+\Delta t$ are given by ${\bf r}_i (t+\Delta t) = {\bf
  r}_i (t) + {\bf v}_i(t) \Delta t$, and $\theta_i(t+\Delta t) =
arg\left[\sum_k \exp(i \theta_k(t)) \right] + \Lambda \xi(i,t)$, where
the summation over $k$ is restricted to those satisfying $|{\bf r}_k -
{\bf r}_i| < R$, and $\xi$ is white noise over the range $(-\pi,
\pi]$. It is known that the system undergoes a transition from an
ordered state to a disordered state as the noise
strength $\Lambda$ is increased \cite{Vicsek}. 

\begin{figure}
\includegraphics[scale = 0.33]{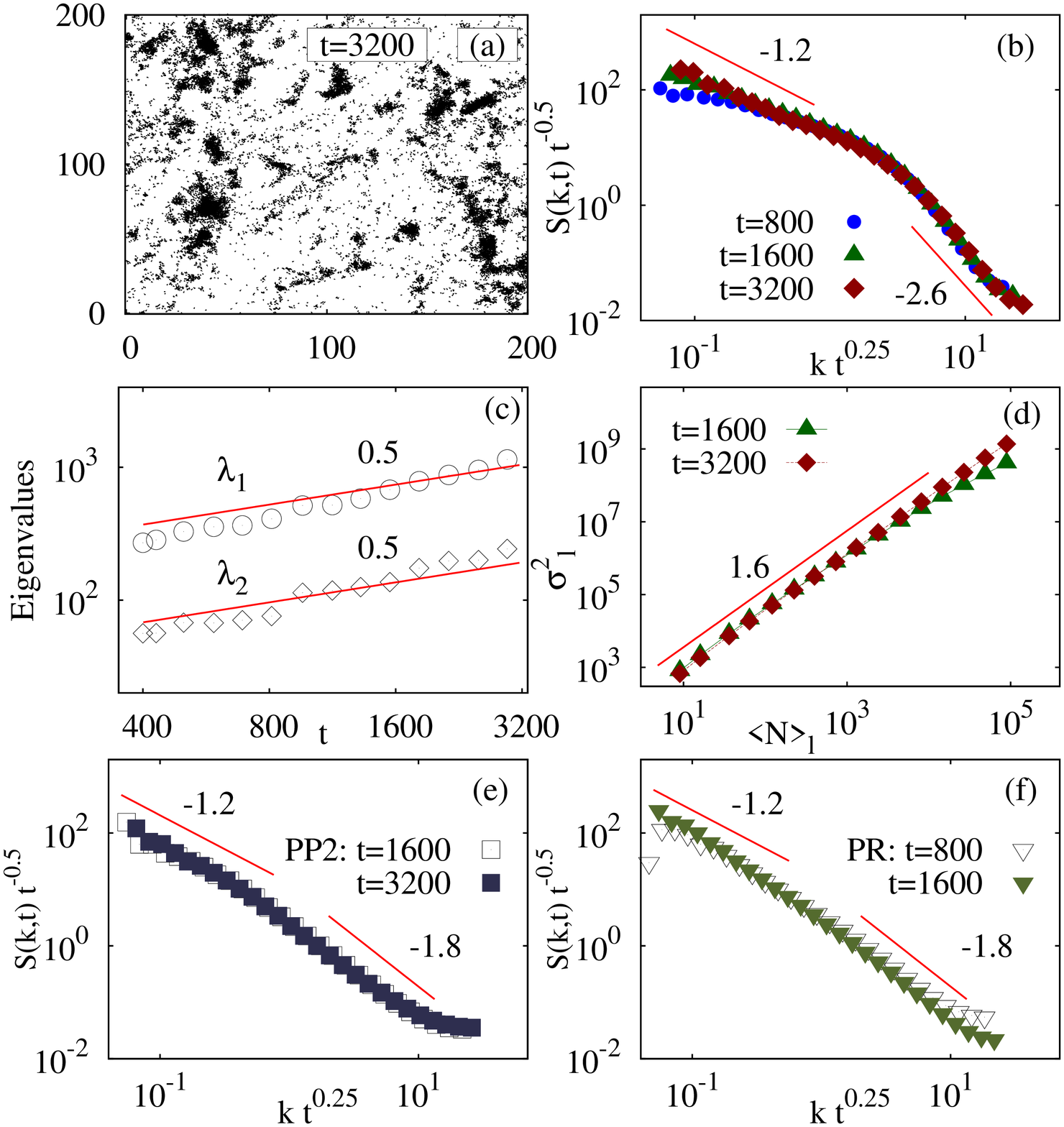}
\caption{(a)--(d): Simulation results for the PP(1) model (system 
size $1024 \times 1024$, $\rho = 1.0$, $v_0=0.5$, and $\Lambda = 0.3$). 
(a) Snapshot of a part of the system at $t=3200$ showing the domain structure. 
(b) Plot of scaled structure function decays as a power law with
exponents $-2.6$ [large $k\mathcal{L}(t)$] and $-1.2$ [small
$k\mathcal{L}(t)$].
(c) The eigenvalues of the inertia tensor for the largest cluster 
grow as $\sim t^{0.5}$.
(d) Variance of number $\sigma_l^2$ $\sim$ $\langle N \rangle ^{1.6}$.
(e) PP(2) model (parameters same as PP(1)):  scaled structure function 
decays as a power law with exponents $-1.8$ [large $k\mathcal{L}(t)$] and
 $-1.2$ [small $k\mathcal{L}(t)$]. 
(f) PR model (system size $1024 \times 1024$, $\rho = 1.0$, $v_0=0.5$, 
and $\Lambda = 0.2$):  scaled structure function 
decays as a power law with exponents $-1.8$ [large $k\mathcal{L}(t)$] and
 $-1.2$ [small $k\mathcal{L}(t)$]. 
}
\label{vicsek}
\end{figure}

We choose parameter values for which the steady state is polar ordered
with no density bands, and study numerically the density structures in
the coarsening regime--- a typical snapshot of the density clusters in shown in
\fref{vicsek}(a).  In the time regime studied, $C({\bf r},t)$ has
no directional anisotropy. Hydrodynamic theory predicts a length scale
$\mathcal{L}(t) \sim t^{5/6}$ \cite{TonarPrl98,TonarRev}.
Interestingly, we find that $C({\bf r},t)$ and the corresponding
scaled structure function $S(k,t)/{\mathcal{L}}^2$ [see \fref{vicsek}(b)] 
shows a data collapse for a completely different
coarsening length $\mathcal{L}(t) \sim t^{\gamma}$ with $\gamma = 0.25
\pm 0.05$. To understand the physical origin of this length scale, we
studied the two eigenvalues $\lambda_1$ and $\lambda_2$ of the inertia
tensor of the largest cluster. Both of these grow as $\sim t^{0.5}$
[see \fref{vicsek}(c)], implying that the radii of the large clusters 
grow as $t^{0.5}$, determining the length scale $\mathcal{L}(t)$.  
The $S(k,t)$ [\fref{vicsek}(b)] consists
of two distinct power laws with exponents $-1.2$ for small
$k\mathcal{L}(t)$ and $-2.6 \pm 0.1$ for large $k\mathcal{L}(t)$; the
former has been known in hydrodynamic theory \cite{TonarPrl98}, but we
highlight the latter, signifying violation of Porod
law.  In real space, the latter implies that $C({\bf r},t)$ has a cusp
of the form $a - b|{\bf r}/\mathcal{L}(t)|^{\beta_1}$ with
$\beta_1=0.6 \pm 0.1$ for $|{\bf r}|/\mathcal{L}(t) \ll 1$, and a
second power law $|{\bf r}|^{-\eta}$ with $\eta=0.8\pm 0.1$ for $|{\bf
r}|/\mathcal{L}(t) \geq 1$.  Due to this crossover, the GNF exponent
$\alpha$, determined from Eq.~(\ref{eq1}), depends only on the
exponent $\eta$: \beq \alpha = 2 - \eta/d.
\label{eq2}
\eeq
In the coarsening regime, from a direct measurement of $\sigma_l^2$, 
we find $\alpha = 1.6$ [see \fref{vicsek}(d)], 
consistent with \eref{eq2}, and 
measurements in the steady state \cite{ChateDisk}.

Interestingly, the structure functions in two other polar models have
the same qualitative behaviour. A modified version of the PP(1) model
was studied in Ref.~\cite{ChateCoh,*ChatePre08}, which we refer to as
PP(2) model. The rules of the PP(2) model are same as those of the
PP(1) model except that the the new positions depend on the new
velocities: ${\bf r}_i (t+\Delta t) = {\bf r}_i (t) + {\bf
  v}_i(t+\Delta t) \Delta t$.  Most macroscopic features remain the
same as PP(1), except that the steady state configurations have
density bands \cite{ChateCoh,*ChatePre08}.  In the time regime that we
study, $C({\bf r},t)$ is isotropic and bands do not form. The
structure function is shown in \fref{vicsek}(e) -- we find, as in
PP(1), $\mathcal{L}(t) \sim t^{0.25}$ and the small $|{\bf
  k}|\mathcal{L}(t)$ behaviour implies $\eta = 0.8$. However, for
large $|{\bf k}|\mathcal{L}(t)$, $S({\bf k},t)$ is  a distinct power
law with exponent $-1.8$, implying a `divergence' (as opposed to a
cusp) of $C({\bf r},t)$ for small $|{\bf r}|/\mathcal{L}(t)$ as
$|{\bf r}/\mathcal{L}(t)|^{-\beta_2}$, with $\beta_2 = 0.2 \pm 0.1$
--- again showing non-Porod behaviour.

Next we studied a PR model defined in in Ref.~\cite{ChateSpr}. The
time evolution of the $\theta_i$'s in the PR model differs from that
in the PP(2) model: $\theta_i(t+\Delta t) = arg\left[\sum_k
\mathrm{sign}\left[ \cos(\theta_k(t)-\theta_i(t))\right] \exp(i
\theta_k(t)) \right] + \Lambda \xi(i,t)$, and $\Lambda \in
(-\pi/2,\pi/2]$.  We find that the scaling of $\mathcal{L}(t)$, as
well as the shape of $S({\bf k},t)$ of the PR model is similar to
the PP(2) model [see \fref{vicsek}(f)].

In summary, the models PP(1), PP(2) and PR share the following common
features: They have a coarsening length scale $\mathcal{L}(t) \sim
t^{0.25}$. At small $|{\bf r}|/ \mathcal{L}(t)$ the correlation
functions violate Porod law either as a cusp or as a power law
divergence. For large $|{\bf r}|/ \mathcal{L}(t)$, they exhibit a
generic second power law decay with exponent $\eta = 0.8$, which
determines the GNF exponent $\alpha = 1.6$.  Thus, for polar models ({\it
  type 1}), the non-Porod behaviour, and the GNF, characterize
distinct sources of fluctuation.

\begin{figure}
\includegraphics[scale = 0.33]{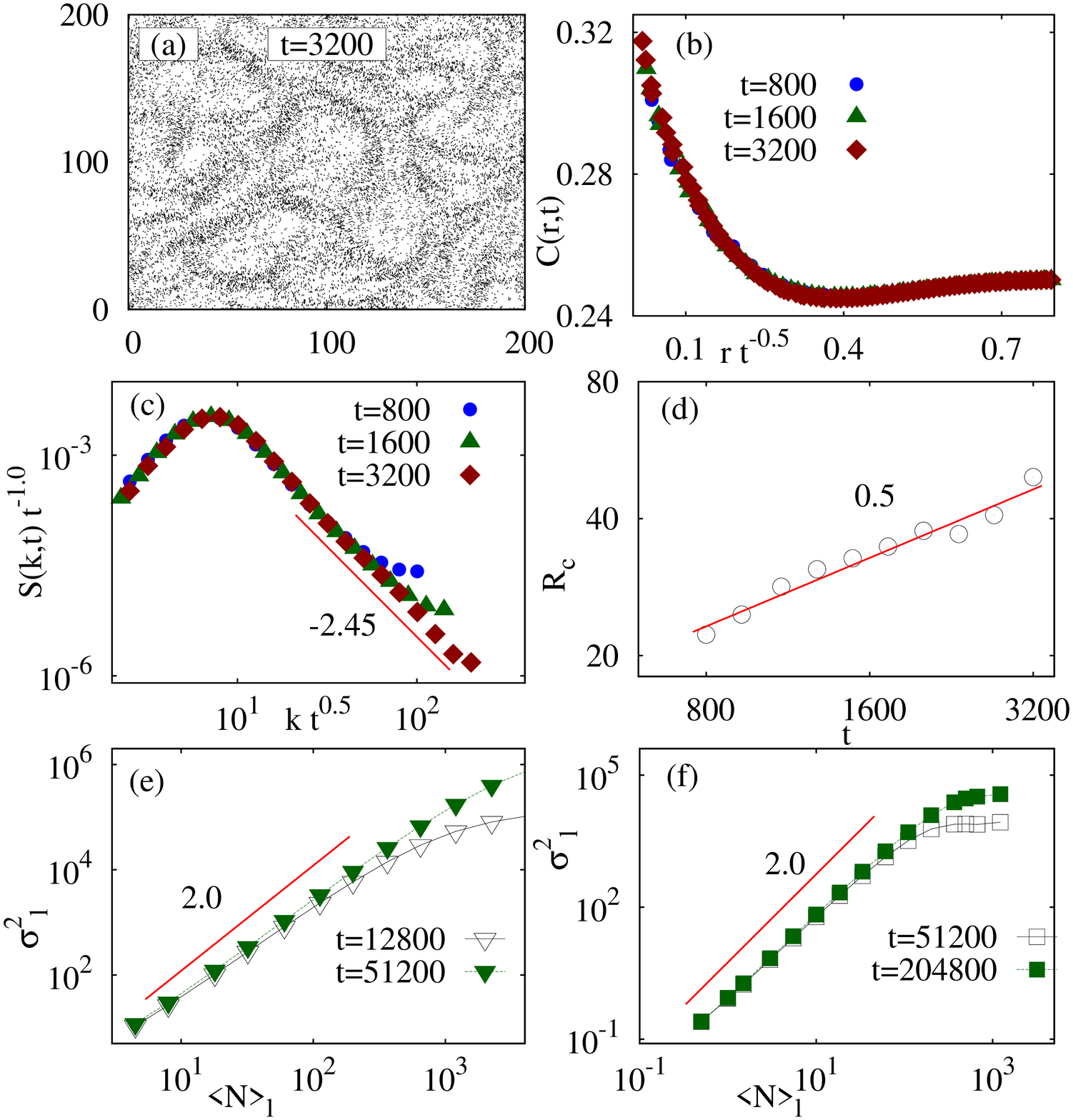}
\caption{(a)--(e): Simulation results for the AR model (system 
size $1024 \times 1024$, $\rho = 0.5$, $v_0=0.3$, and $\Lambda = 0.08$).
(a) Snapshot of a part of the system at $t=3200$ showing the domain structure. 
(b) $C(r,t)$ versus $r/\mathcal{L}(t)$ showing data collapse.   
(c) The scaled structure function is a power law with
exponent $-2.5$ at large $k\mathcal{L}(t)$.
(d) The mean cell radius $R_c(t) \sim \mathcal{L}(t)$.
(e) Number fluctuations $\sigma_l^2$ $\sim$ $\langle N \rangle ^{2.0}$.
(f) Sliding particle model ($10^6$ lattice sites, particle density $0.5$): 
$\sigma_l^2$ $\sim$ $\langle N \rangle ^{2.0}$.
}
\label{Rama}
\end{figure}

{\it Type 2}: We study the discrete AR model introduced in
Ref.~\cite{ChateAnPrl06}. The $\theta_i$'s now evolve as follows.  The
traceless two dimensional matrix $Q_{jk}= \langle v_j v_k \rangle -
\frac{1}{2} \delta_{jk}$ (with $v_j$ denoting the components of the
unit velocity vectors) is calculated, where the average is done over
the particles that are in the disk of radius $R$, centered about
particle $i$.  If $\bar{\Theta}$ denotes the direction of the largest
eigenvector of $Q$, then $\theta_i(t+\Delta t) = \bar{\Theta} +
\Lambda \xi(i,t)$, with $\Lambda \in (-\pi/2,\pi/2]$. The positions
${\bf r}_i (t+\Delta t) = {\bf r}_i (t) \pm {\bf v}_i(t+\Delta t)
\Delta t$, with the signs being chosen randomly with equal
probability. The steady state of the AR model is characterized by
nematically ordered bands, however, in the coarsening regime that we
study they do not arise. Rather very interesting cell-like
structures--- low density zones with high density contours--- form
[see \fref{Rama}(a)], whose radii increase with time.  The data
for $C({\bf r},t)$ and $S({\bf k},t)/{\mathcal{L}}^2$ for different
times, collapse when plotted against $|{\bf r}|/\mathcal{L}(t)$ or
$|{\bf k}|\mathcal{L}(t)$, with $\mathcal{L}(t) \sim t^{0.5}$ [see
\fref{Rama}(b) and (c)]. 
We make an independent estimate of $\mathcal{L}(t)$ by counting the number of
cell-like zones, thus measuring the mean cell radius $R_c(t)$. 
We obtain $\mathcal{L}(t) \sim R_c \sim t^{0.5}$ [see \fref{Rama}(d)].  For
$|{\bf r}|/\mathcal{L}(t) \ll 1$, $C({\bf r},t) \sim a - b|{\bf
r}/\mathcal{L}(t)|^{\beta_1}$ shows Porod law violation
with a cusp singularity $\beta_1 =
0.45 \pm 0.05$ determined from $S \sim {k\mathcal{L}}^{-2.45}$ 
[see \fref{Rama}(c)].
Unlike the PP models, there is no second power law regime in $C({\bf
r},t)$ for $ |{\bf r}|/\mathcal{L}(t) \gg 1$. The above cusp singularity
of $C({\bf r},t)$ is similar to another discrete model of active
nematics in two dimensions (see Fig.~5 of Ref.~\cite{SriramPrl06}), and a
completely different model of particles sliding under gravity on a
fluctuating interface in one dimension (see Fig.~2 of
Ref.~\cite{DibyenduPrl00}). Due to the similar functional 
form  of $C({\bf r},t)$,  the number fluctuation from \eref{eq1} is:  
\bea \sigma_l^2 \sim |a| \la N \ra^2 - \frac{|b|}{{\mathcal L}^{\beta_1}} 
\la N \ra^{\frac{\beta_1}{d} + 2} + \cdots
\label{GNF_AR}
\eea 
The leading order behaviour $\sigma_l^2 \approx \la N \ra^2$ is
seen in our data for the AR model [see \fref{Rama}(e)], as well as for the
sliding particle system [see \fref{Rama}(f)]. Thus, the scaling $\sigma_l^2
\sim \la N \ra^2$ may arise in systems other than active nematics.
Even in an ordinary density phase segregating system with density
$1/2$ consisting of domains of equal length $\mathcal{L}(t)$, the
correlation function is $1 - |r|/\mathcal{L}(t)$
(satisfying Porod law) for $|r| < 2 \mathcal{L}(t)$, the number
fluctuation is exactly $\sigma_l^2 = \la N \ra^2 - 4 \la N \ra^3/\mathcal{L}(t)
\approx \la N \ra^2$ for large ${\mathcal L}(t)$. 

We make two important observations related to
Eq.~(\ref{GNF_AR}): (i) First, the data in \fref{Rama}(e) and (f),
and also in the experiment of vibrated granular rods \cite{Sriram},
show a visible deviation from the leading $\sigma_l^2 = \la N \ra^2$
behavior. We claim that the subleading term in Eq.~(\ref{GNF_AR}) may
account for this deviation.  We confirmed that $- \sigma_l^2/\la N
\ra^2 + |a|$ indeed scale as $\la N \ra^{0.23}$ (consistent with
$\beta_1 = 0.45$, $d=2$) and $\la N \ra^{0.5}$ (consistent with
$\beta_1 = 0.5$, $d=1$), respectively, for the data in
\fref{Rama}(e) and (f). More remarkably, we fitted the published
experimental data \cite{Sriram} in this way, and concluded that the
experimental system has a $\beta_1 \approx 0.5$ --- that is Porod
law is indeed violated and the exponent is close to the AR model
above. We, thus, suggest that Eq.~(\ref{GNF_AR}) opens up a new
possibility for experimentalists --- by measuring the sub-leading
corrections to GNF, they can indirectly measure Porod law violation.
(ii) Second, we do not see a power law $\sim |{\bf k}|^{-2}$ at
small ${\bf k}$ for $S({\bf k})$ [see \fref{Rama}(c)], as suggested
by continuum theory \cite{SriramEPL03}. The same is true for the
sliding particle model \cite{DibyenduPrl00,*DibyenduPre01}, as well
as the clean phase separating system discussed above --- for the latter
$S({\bf k},t) = 4 \mathcal{L}(t) \sin^{2}
(k\mathcal{L}(t)/2)/(k\mathcal{L}(t))^2$. Instead, these
three examples have a divergence $\sim {\mathcal L}^d$ as ${\bf k}
\rightarrow 0$.  This indicates that $\sigma_l^2 \sim l^d S({\bf k}
\rightarrow 0) \sim l^d/|{\bf k}|^d \sim l^{2d} \sim \la N \ra^2$. 
\begin{figure}
\includegraphics[scale = 0.33]{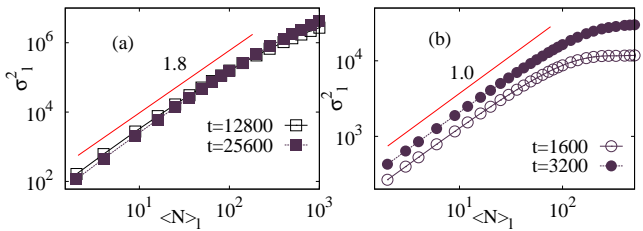}
\caption{Number fluctuations: (a) the granular gas ($L=10^5$, $\rho=1.0$, 
$r_0=0.5$, and $\delta=0.008$) showing GNF with exponent $\alpha=1.8$. 
(b) ballistic aggregation ($L=10^5$, and $\rho=1.0$) shows normal fluctuation 
 with exponent $\alpha=1.0$.
}
\label{type3-4}
\end{figure}

{\it Type 3}: A situation different from type 2  would arise if $C({\bf r},t)$
has a power law form $\sim |{\bf r}/\mathcal{L}(t)|^{-\eta}$ over    
small through large $|{\bf r}/\mathcal{L}(t)|$. On one hand 
there will be non-Porod behaviour, and on the other, the same
exponent contributes to the GNF exponent $\alpha$ and is given by \eref{eq2},
provided $\eta<d$.
Although we are not aware of an active matter system exhibiting such
behaviour, another nonequilibrium system namely a freely cooling
granular gas in one-dimension \cite{MahendraPrl,*MahendraPre11} serves
as an example of this type. Its structure function shows that the
$|{\bf k}|$ space exponent is $-0.8$ \cite{MahendraPrl}, and hence
$\eta = 0.2$. We revisit this model, and calculate the
$\sigma_l^2(t)$ in the coarsening regime. The result is shown in
\fref{type3-4}(a). We find a new GNF exponent value 
$\alpha = 1.8$, consistent with \eref{eq2}.

{\it Type 4}: Central limit theorem as in $\alpha=1$, may hold in 
an interesting situation. This is when dense clusters (with masses
scaling as ${\mathcal L}(t)^d$) appear in isolated locations, leading to
temporal ``intermittency''. In this case, $C({\bf r},t) = \mathcal{L}(t)^d  
\delta({\bf r}) + f(|{\bf r}/{\mathcal L}(t)|)$. 
Due to the presence of the $\delta$-function, the 
form of the scaling function $f(|{\bf r}/{\mathcal
L}(t)|)$ is irrelevant in the calculation of $\sigma_l^2$ from \eref{eq1}. 
Thus, $\alpha = 1$. 
Such situations arise in aggregation models: diffusive or ballistic, wherein
particles aggregate on contact conserving mass. 
The density-density correlation
function for the ballistic system has been studied in molecular dynamics and also
for an equivalent lattice model \cite{SupravatEpl,*MahendraPre}. For
$|{\bf r}/{\mathcal L}(t)| > 0$, the scaling form of the correlation
function starts from a low value, rises linearly and then saturates, 
with increasing $|{\bf r}/{\mathcal L}(t)|$. While Porod law holds,
$C({\bf r},t)$ also has a term $\mathcal{L}(t) 
\delta({\bf r})$. We measure $\sigma_l^2$ in simulations of the lattice version
of the model, and clearly see $\alpha = 1$ as predicted
(see Fig. \ref{type3-4}(b)).
We note that the {\it type 4} is distinct from the other types
in another respect: $\la N^k \ra \propto \la N \ra$,
for all integer $k \geq 2$, a consequence of statistics being dominated by the
largest cluster. We check that this is true for ballistic aggregation. 
For all other cases ({\it types 1-3}) $\la N^k \ra \propto
\la N \ra^{\gamma k}$, where the exponent $\gamma$ is specific to a system.

In summary, we studied well known discrete models of active matter and
some other nonequilibrium systems, to understand similarities and
differences in their density structures.  All the active matter
systems that we studied are shown to violate Porod law, and the
non-Porod behaviour is quantified by various new exponents.  We
categorised the relationship between spatial density-density
correlation function and giant number fluctuation into four classes.
The first one, formed by polar particles and rods, shows two distinct
scaling behaviour at small and large length scales. These models were
shown to have a new coarsening length scale $\mathcal{L}(t) \sim
t^{0.25}$.  The non-Porod behaviour does not influence GNF.  In the
second class, examples being apolar rods or particles sliding on
fluctuating surfaces, we showed that the subleading corrections to the
number fluctuations may help experimentalists to detect Porod law
violation. Also the known discrete models belonging to this class
exhibit GNF for a different reason than proposed by continuum theory
of active nematics.  In the third class, the scaled correlation
function exhibits power law divergence at small scale. An example for
this class is the one-dimensional freely cooling granular gas, for
which we found a new GNF exponent $\alpha = 1.8$.  Finally,
aggregation models are examples of a fourth distinct class, for which
the number fluctuations satisfy central limit theorem.  We hope that
this study will encourage experimentalists to probe density structures
in detail in the future.



\ifx\mcitethebibliography\mciteundefinedmacro
\PackageError{unsrtM.bst}{mciteplus.sty has not been loaded}
{This bibstyle requires the use of the mciteplus package.}\fi

\end{document}